\documentclass[aps,prl,twocolumn,showpacs,superscriptaddress,groupedaddress]{revtex4}

\usepackage{graphicx}
\usepackage{color}
\usepackage{dcolumn}
\usepackage{bm}
\usepackage{amssymb}
\usepackage{amsmath}
\usepackage{textcomp}

\begin{document}

\title{Fast magnetic field annihilation in the relativistic collisionless regime driven
by two ultra-short high-intensity laser pulses}

\author{Y. J. Gu}
\email{yanjun.gu@eli-beams.eu}
\affiliation{Institute of Physics of the ASCR, ELI-Beamlines, Na Slovance 2, 18221 Prague, Czech Republic}
\author{O. Klimo}
\affiliation{Institute of Physics of the ASCR, ELI-Beamlines, Na Slovance 2, 18221 Prague, Czech Republic}
\affiliation{FNSPE, Czech Technical University in Prague, 11519 Prague, Czech Republic}
\author{D. Kumar}
\affiliation{Institute of Physics of the ASCR, ELI-Beamlines, Na Slovance 2, 18221 Prague, Czech Republic}
\author{Y. Liu}
\affiliation{Institute of Physics of the ASCR, ELI-Beamlines, Na Slovance 2, 18221 Prague, Czech Republic}
\author{S. K. Singh}
\affiliation{Institute of Physics of the ASCR, ELI-Beamlines, Na Slovance 2, 18221 Prague, Czech Republic}
\author{T. Zh. Esirkepov}
\affiliation{
Kansai Photon Science Institute, Japan Atomic Energy Agency,
8-1-7 Umemidai, Kizugawa-shi, Kyoto, 619-0215 Japan}
\author{S. V. Bulanov}
\affiliation{
Kansai Photon Science Institute, Japan Atomic Energy Agency,
8-1-7 Umemidai, Kizugawa-shi, Kyoto, 619-0215 Japan}
\author{S. Weber}
\affiliation{Institute of Physics of the ASCR, ELI-Beamlines, Na Slovance 2, 18221 Prague, Czech Republic}
\author{G. Korn}
\affiliation{Institute of Physics of the ASCR, ELI-Beamlines, Na Slovance 2, 18221 Prague, Czech Republic}

\date{\today}

\begin{abstract}
The
magnetic quadrupole structure formation
during the interaction of two ultra-short high power laser pulses
with a collisionless plasma
is demonstrated with 2.5-dimensional particle-in-cell simulations.
The subsequent expansion of the quadrupole is accompanied by magnetic field annihilation
in the ultrarelativistic regime, when
the magnetic field can not be
sustained by the plasma current. This results
in a dominant contribution of the displacement current
exciting a strong large scale electric field.
This field
leads to the conversion of magnetic energy into kinetic energy of accelerated electrons
inside the thin current sheet.

\end{abstract}


\maketitle

\
Magnetic reconnection is a process which changes the topology of the magnetic field lines and allows the transfer of magnetic field energy into charged particles energy \cite{BiskampMR}.
It plays a fundamental role in the dynamics of magnetic confinement
thermonuclear  plasma  \cite{RW} and has been
considered as a plausible mechanism for high energy charged particle
generation in space plasmas \cite{Giovanelly,  Dungey, CRA, Birn}.
This process is accompanied by the current sheet formation
\cite{Parker, Sweet, Syrovatskii1971}, where the oppositely directed magnetic fields annihilate.
Annihilation is a basic mechanism for reconnection and has been investigated within the framework of dissipative magnetohydrodynamics\cite{IP}. In the limit of ultrarelativistic plasma dynamics
the magnetic field annihilation acquires principally different properties related to the existence
of a limiting value of the electron density \cite{Syrovatskii1966}.
This is due to the relativistic constraint on the upper limit of the particle
velocity which can not exceed the speed of light
in vacuum and can sustain only a limiting magnetic field strength.

The development of high power laser technology \cite{GMCPA} allows to access new regimes of magnetic field annihilation. When a high intensity laser pulse interacts with a plasma target the accelerated electron bunches generate strong regular magnetic fields\cite{Forslund, Askaryan1994}. The self-generated magnetic field in inhomogeneous near critical density plasma also enhances fast ion generation\cite{MagVort}. It has been predicted that nontrivial topology of self-generated magnetic field configurations should invariably lead to magnetic reconnection\cite{Askaryan1995}. Recent experiments have observed plasma outflows with MeV electrons and plasmoids generated in the reconnection current sheet\cite{Nilson2006, Zhong2010}. These experiments used long nano-second laser pulses and investigated reconnection in a parameter regime where the conditions for fast relativistic magnetic field line reconnection can not be reached.

In this letter, a fast magnetic field annihilation in the relativistic collisionless
regime driven by two  parallel synchronized laser pulses with ultra-high intensity
and femto-second pulse duration is studied with kinetic simulations. Different from Ref. \cite{Ping2014}, we present the magnetic annihilation in the rear of the target where the plasma density is very low. The convection current is negligible in the annihilation region and the variation of the magnetic field is compensated by the displacement current. The inductive electric field propagates forward and accelerates electrons and positrons in the current sheet.

\begin{figure}
\begin{center}
\resizebox{65mm}{!}{\includegraphics{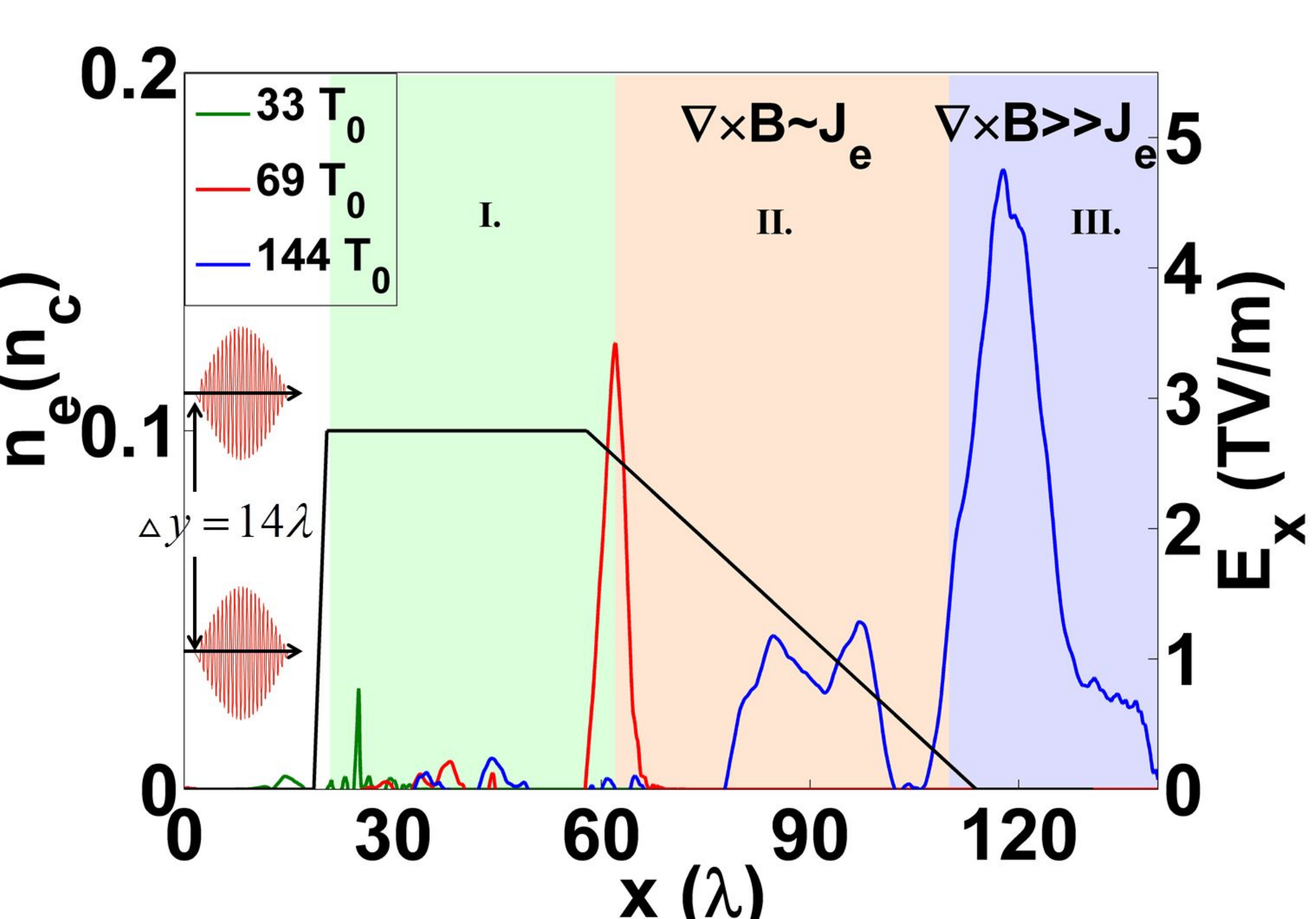}}
\caption{(color online)
Setup of the simulation. The longitudinal electric field values along $y=0$ are taken at
 $33~T_0$ (green), $69~T_0$ (red) and $144~T_0$ (blue), where $T_0$ is the laser period. Zones I, II and III correspond
to the regions of magnetic field generation, expansion and annihilation, respectively.}
\label{Fig1}
\end{center}
\end{figure}

The Particle-In-Cell (PIC) simulations are performed with the relativistic electromagnetic
code EPOCH \cite{EPOCH2014}. Two s-polarized Gaussian pulses with peak
intensity of $\mathrm{10^{21}~W/cm^2}$ are incident in the $x$ direction and focused on
the left edge of the target.
By choosing s-polarization, the effects of the the high frequency laser field are mitigated,
which is helpful for observing clearly
the self-generated magnetic field and the inductive electric field.
The pulses durations are $\tau=15~\mathrm{fs}$ and the laser
spot sizes (FWHM) are $3~\lambda$, where $\lambda=2\pi c/\omega=1~\mu m$ is the laser wavelength
and $\omega$ is the laser frequency.
The optical axes of the two laser pulses are at $y=\pm7~\lambda$.
The separation guarantees the formation of two independent electron bubbles,
which do not overlap with each other.
The simulation box is $195~\lambda \times 340~\lambda$ in $x$ and $y$, respectively. The transverse size is large enough to avoid boundary effects.
The preformed hydrogen plasma is located in $20~\lambda<x<122~\lambda$
with non-uniform density distribution in the $x$ direction and uniform density in the $y$ direction.
The longitudinal density profile is shown in Fig. \ref{Fig1}.
The plasma density linearly
increases from $0$ to $0.1~n_c$ in the interval $20~\lambda<x<22~\lambda$,
and then remains constant in $22~\lambda<x<62~\lambda$. For $62~\lambda<x<122~\lambda$,
there is a downramp region, where the density linearly decreases to zero.
Here $n_c=m_e \varepsilon _0 \omega ^2 /e^2$ is the plasma critical density,
$m_e$, $e$ and $\varepsilon _0$ represent electron mass, electric charge, and
vacuum permittivity, respectively. The transverse size of the target is $40~\lambda$.
The mesh size is $\delta x=\delta y=\lambda /20$.
There are $10^7$ pseudo-particles in the simulation box and all the particles are
initially cold. Open boundary conditions are employed for both particles and fields.

\begin{figure}
\resizebox{42mm}{!}{\includegraphics{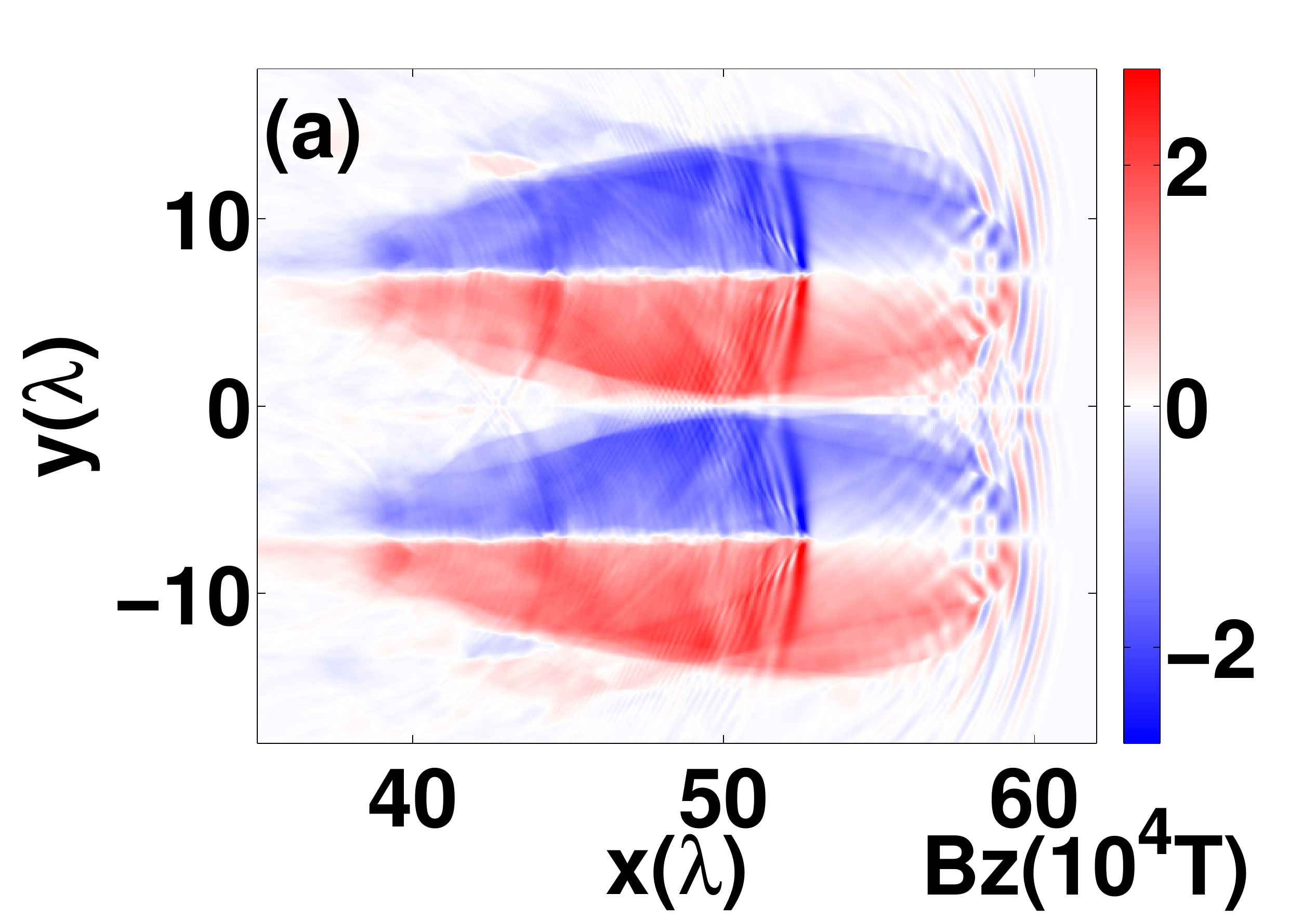}}
\resizebox{42mm}{!}{\includegraphics{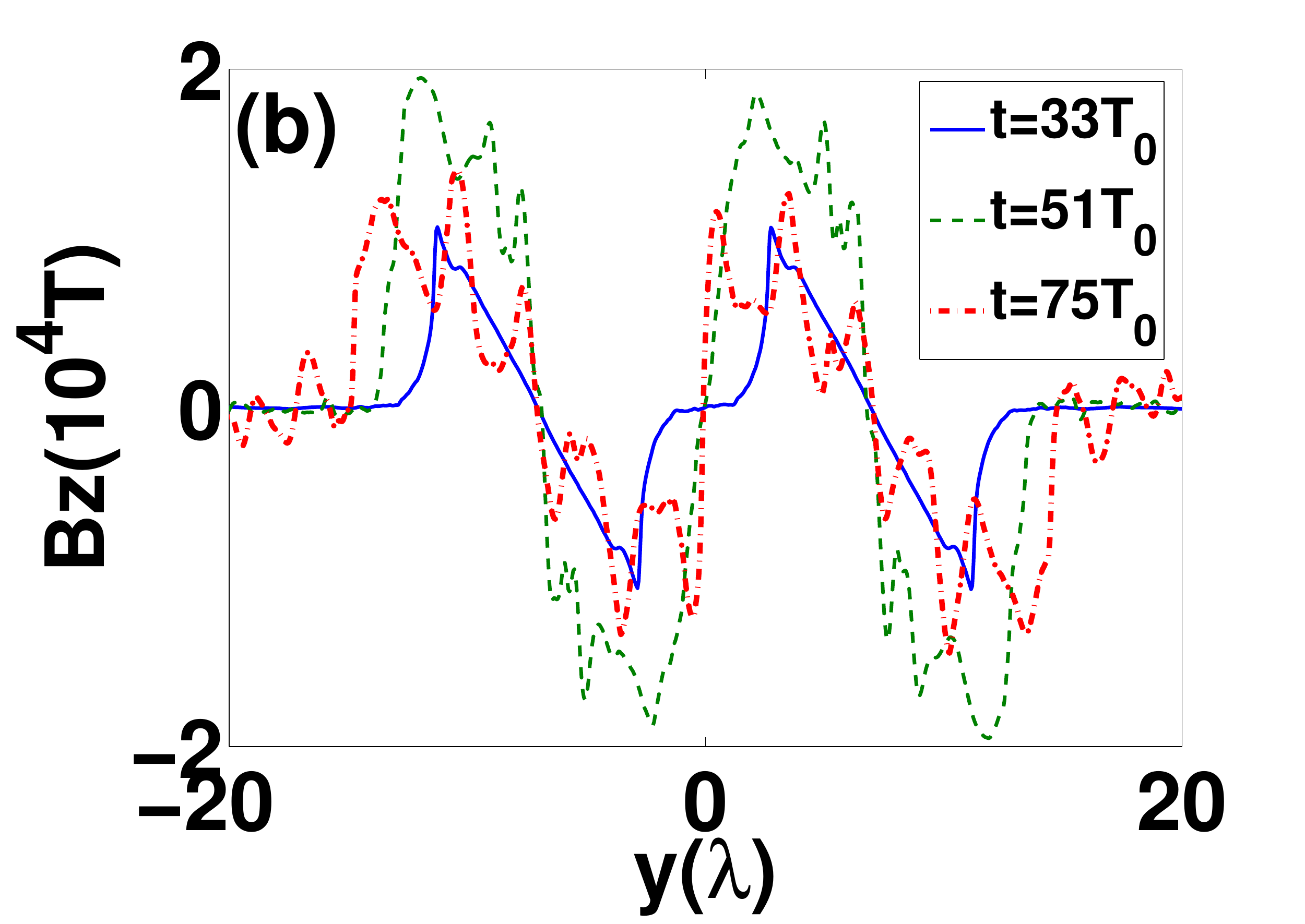}}
\caption{(color online)
(a) The z-component of the magnetic field ($B_z$) distribution at $t=63~T_0$.
(b) The expansion of the magnetic dipole structures.
The blue solid line, green dashed line and red dotted line represent
the $B_z$ profile along $x=27~\lambda$, $40~\lambda$ and
$68~\lambda$ at $33~T_0$, $51~T_0$ and $75~T_0$, respectively.}
\label{Fig2}
\end{figure}

The propagation of the laser pulses inside the plasma results in
the formation of  two electron bubbles due to the laser wakefield effect (\cite{ESL} and references cited therein).
The strong wakefields accelerate
electrons in the longitudinal direction. According to Ampere's law, the electric currents produce
magnetic fields, which is $B_z$ in the 2D case.
As a result, a magnetic quadrupole configuration is formed as shown in Fig. \ref{Fig2}(a).
At this moment, the two magnetic dipoles do not touch each other in the vicinity of the
central axis $(y=0)$. The maximum magnetic field can be calculated by using Faraday's
law as
$\nabla \times \mathbf{B}=\mu _0 \mathbf{J_e}+ \mu _0\varepsilon _0 \partial \mathbf{E}/ \partial t$
with $\mu_0$ being the vacuum permeability.
Assuming the quasistatic conditions, we take
the magnetic dipole transverse size to be of the order of $R \approx c\sqrt{a_0} /\omega_p$,
which corresponds to the radius of the self-focusing filament for the laser pulse with the
normalized amplitude $a_0 =eE/m_e\omega c$, where the Langmuir frequency
is $\omega_p=(4\pi n e^2/m_e)^{1/2}$. Then the magnetic field can be estimated as
$B_z =\mu _0 n_0 ec^2\sqrt{\gamma} /\omega _p$ with the Lorentz factor $\gamma \approx a_0$  \cite{Askaryan1994}. It gives
$B_z \approx 1.8 \times 10^4~\rm{T}$ in our case,
which is consistent with the simulation result.

When the bubbles propagate into
the density downramp region, where $n(x)=-n_0 (x-122~\lambda)/(60~\lambda)$,
both the bubble size and the magnetic field dipoles expand transversely
since they experience forces acting on the vortex proportional to $\nabla n \times \mathbf{\Omega}$ \cite{Nycander1990}. Here $\mathbf{\Omega}$
is the potential vorticity. Figure 2(b)
shows the profile of $B_z$ along $x=27~\lambda$, $40~\lambda$ and
$68~\lambda$ at $33~T_0$, $51~T_0$ and $75~T_0$, respectively.
During the time between $33~T_0$ and $51~T_0$, the quadrupole is still propagating
in the uniform density region and the displacement between the blue
solid peak and the green dashed peak is about $0.3~\lambda$.
During this stage the growth of the magnetic field amplitude (about up to 8500~T) is
significant  since more and more electrons are
trapped and accelerated by the wakefields.
The situation changes, when the
quadrupole enters the inhomogeneous region, where the displacement between
the green dashed peak and the red dotted peak becomes about $1.6~\lambda$.
At the same time, the magnetic field maximum amplitude slightly decreases which is
in accordance with Ertel's theorem \cite{Zakharov1997}.
The shift of the peak positions proves that
the magnetic fields are expanding in the transverse direction.
As a result, the two magnetic dipoles approach each other
near the central axis, creating steep gradients in the magnetic fields which in turn facilitate annihilation.

\begin{figure}
\resizebox{42mm}{!}{\includegraphics{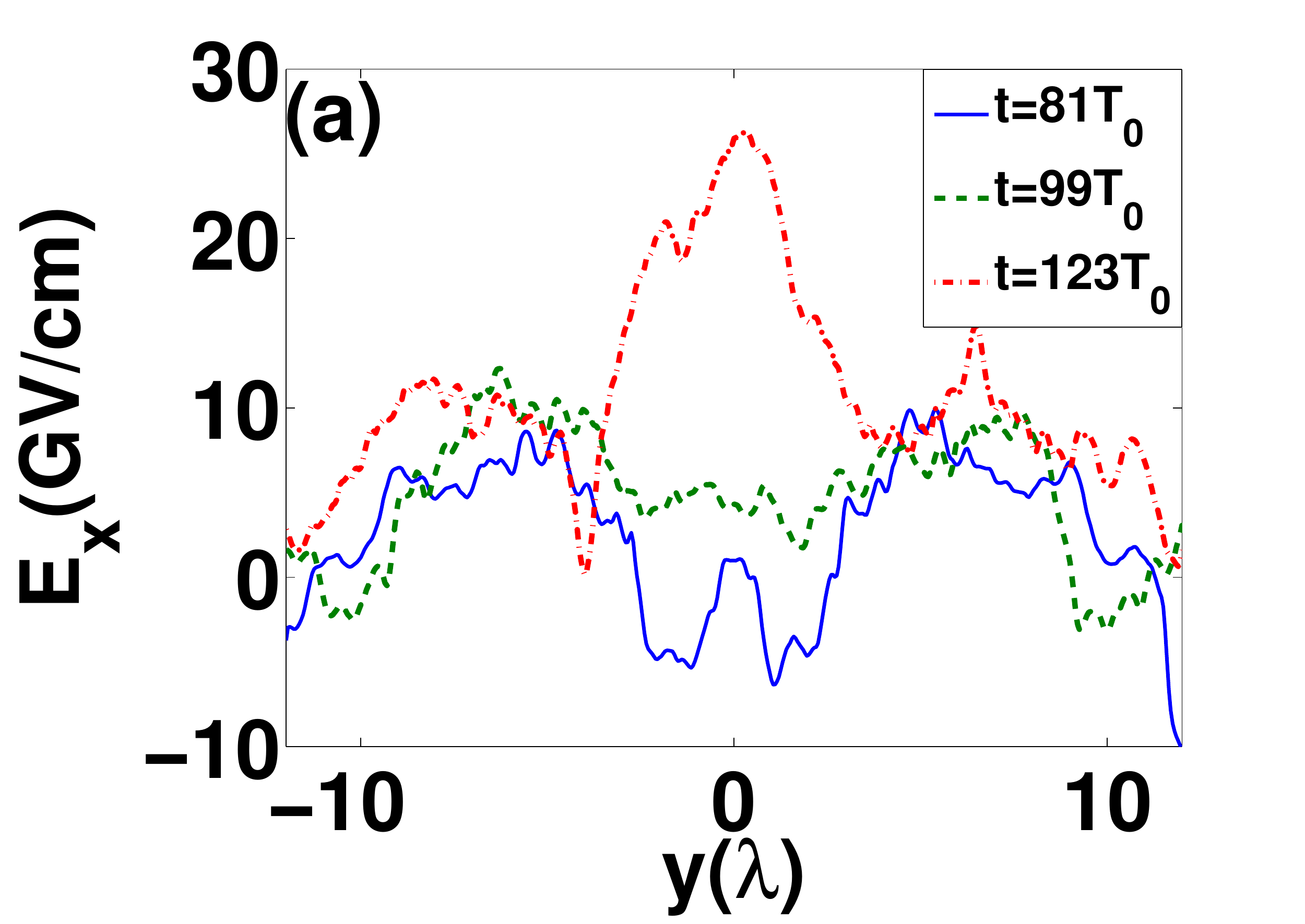}}
\resizebox{42mm}{!}{\includegraphics{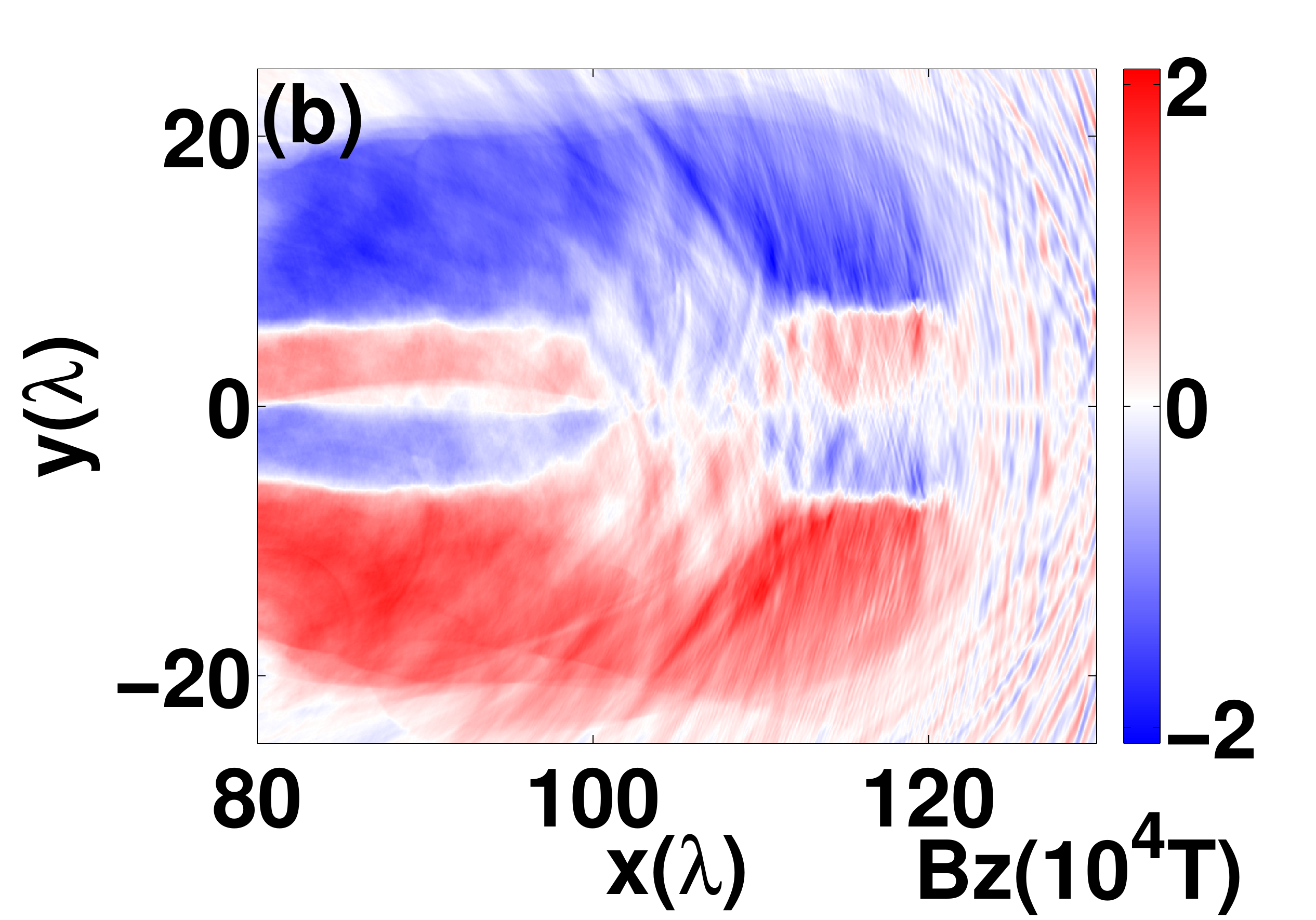}}
\resizebox{42mm}{!}{\includegraphics{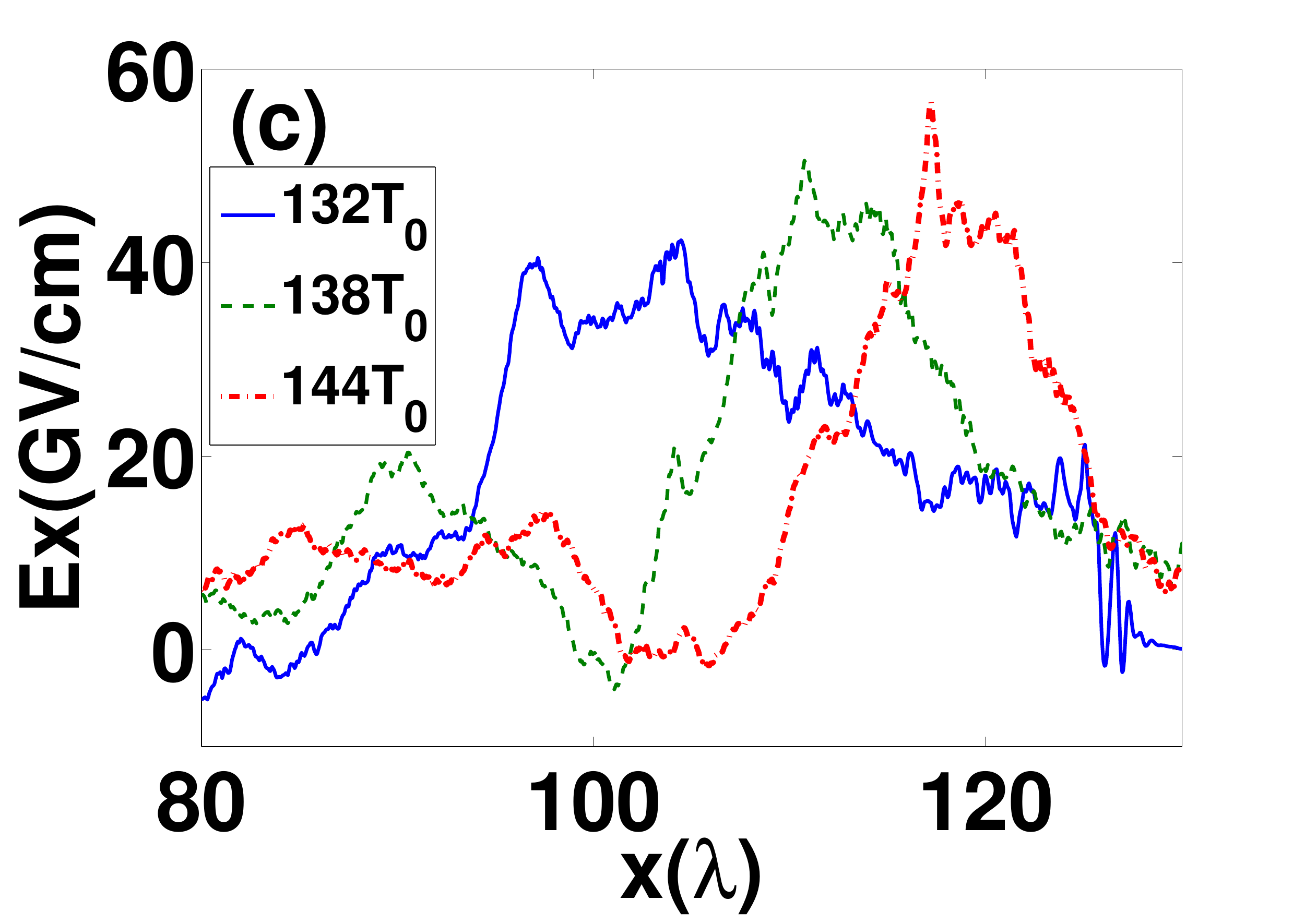}}
\resizebox{42mm}{!}{\includegraphics{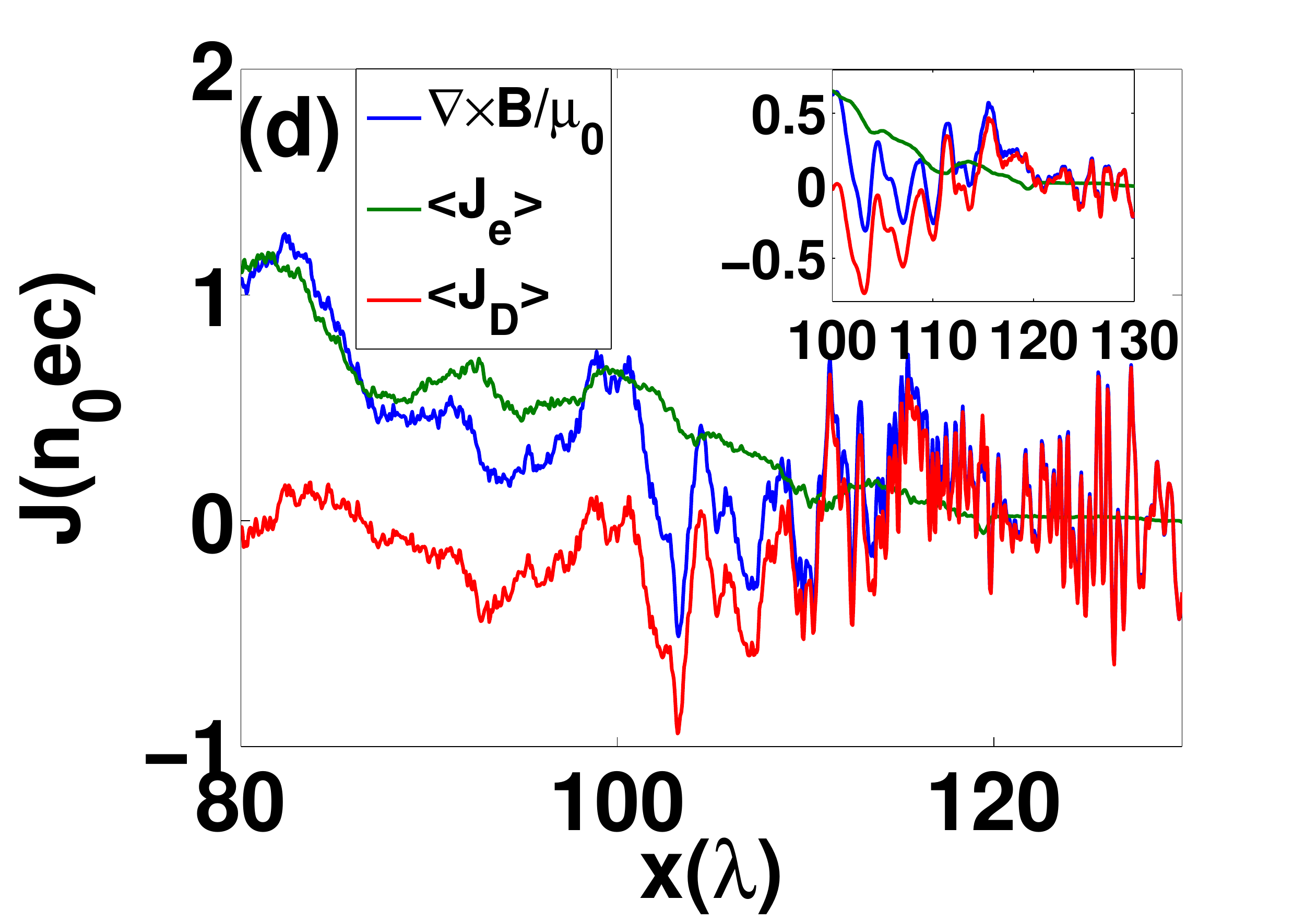}}
\caption{(color online)
(a) The growth of the inductive longitudinal electric field.
The blue solid line, green dashed line and red dotted line represent
the $E_x$ profile along $x=65~\lambda$, $80~\lambda$ and $90~\lambda$
at $t=81~T_0$, $99~T_0$ and $123~T_0$, respectively.
(b) $B_z$ distribution at $138~T_0$.
(c) The profiles of $E_x$ inside the current sheet at $132~T_0$ (blue),
$138~T_0$ (green) and $144~T_0$ (red), respectively.
(d) Contributions of different terms in the Faraday's law at $138~T_0$ along the
$x$-direction and the spatial averaged profile inside the current sheet ($-\lambda < y < \lambda$):
$\frac{1}{\mu_0}<(\nabla \times \mathbf{B})>_x$ (blue),
the convection current density $<\mathbf{J_e}>_x$ (green) and the displacement
current density $<\mathbf{J_D}>_x$ (red). Inset: The smoothed value near the region
where the inductive electric field grows.}
\label{Fig3}
\end{figure}

The central axis becomes the line
where the opposite magnetic field lines annihilate and
rearrange the topology. In the vicinity of the central line,
the current sheet is formed, which is along the $x$ direction.
Magnetic annihilation in the current sheet region is accompanied
by the generation of the
inductive electric field. It is regarded as an important signature
of magnetic reconnection. The longitudinal electric field distribution along
$x=65~\lambda$, $80~\lambda$ and $90~\lambda$ lines at the instants of
$t=81~T_0$, $99~T_0$ and $123~T_0$ respectively
are plotted in Fig. \ref{Fig3}(a). At $81~T_0$ and $99~T_0$,
the intense electric field is located near the laser axes ($y=\pm 7~\lambda$).
Here the electric field corresponds to the plasma wave:
$E_0 =cm_e \omega _p /e \approx 10~\mathrm{GV/cm}$. However,
the longitudinal electric field in the current sheet (near $(y=0)$)
becomes much higher at $t=123~T_0$, while the magnitude of the
electric field along laser axes still keeps the same order.
The extra contribution comes from the magnetic annihilation which releases
the magnetic energy converting it into electric fields. The magnetic field distribution
at $138~T_0$ is shown in Fig. 3(b). We note that a region in the current
sheet around $100~\lambda <x<110~\lambda$, where the
magnetic quadrupole breaks, has unique properties.
Behind the breaking region, the magnetic
field is quite smooth with continuous distribution.
However, in front of the breaking region, the magnetic field becomes filamented and
disrupted, which indicates that the magnetic field lines with
opposite directions are annihilating.
As a result, this filamented region also corresponds to the region of location of
strong inductive longitudinal electric field. This is well seen in
the $E_x$ profiles in Fig. 3(c). The green line shows the
same moment with (b) and has the peak ($E_x > 40~\mathrm{GV/cm}$)
around $110~\lambda <x<115~\lambda$. Fig. 3(c) also shows
the evolution of the longitudinal electric field in the current sheet.
The strong inductive electric field moves forward almost with the speed
of light.

To investigate the regime of the inductive electric field growth,
we compare the contributions of different terms in Faraday's law.
The profiles of $(1/\mu_0)<(\nabla \times \mathbf{B})>_x$,
the convection electric current density $<\mathbf{J_e}>_x=-en {\bf v}_x$ and the displacement
current $<\mathbf{J_D}>_x=\varepsilon _0 \partial_t E_x$ are shown in Fig. 3(d).
In the region $80~\lambda <x< 110~\lambda$, there are return electrons which induce a strong convection
current. Here $(\nabla \times \mathbf{B})_x$ is balanced by $<\mathbf{J_e}>_x$. However,
in the region of $x>110~\lambda$, the electron density is very low due to
the downramp distribution. The variation of magnetic fields can no longer be compensated by the convection current. Therefore, $(\nabla \times \mathbf{B})_x$ is balanced by the displacement current. This induces the
growth of the inductive electric field. The high frequency
oscillations of $<\mathbf{J_D}>_x$ also agree well with the
filaments in magnetic field distribution. After smoothing
the oscillation (see Fig. 3(d), inset), $(\nabla \times \mathbf{B})_x$
becomes precisely in accordance with the displacement current within
the region of the strong inductive electric field.

The electrons within the current sheet get extra energy compared to the electrons
 outside of the current layer due to the contribution of the inductive electric field. Figures 4(a) and (b) show the
distributions of electron longitudinal momentum $p_x$ at $69~T_0$ and $138~T_0$,
which represent the situation without and with magnetic annihilation effects, respectively.
The momenta of the electrons within the current sheet ($-2~\lambda<y<2~\lambda$)
and that of the electrons near the outer wings of laser axis ($8~\lambda< y<25~\lambda$
and $-25~\lambda< y<-8~\lambda$) are comparable in magnitude at $t=69~T_0$. The contributions to $p_x$ at this moment come from the electron
oscillations in the plasma waves and from the return electrons along the bubble shell.
However, at $t=138~T_0$ with the process of magnetic annihilation, a strong backward
accelerated electron bunch is formed in the current sheet. The maximum momentum
reaches $-110~m_e c$, while the momentum growth of the electrons in the wings is not
significant. Furthermore, the origin of the backward accelerated electrons in the
current corresponding to the action of the strong displacement current, \textit{i.e.} to the
inductive electric field as is shown in Figure 3. This also demonstrates that the electron
acceleration is driven by the magnetic annihilation.
Such a kind of violent electron acceleration in the current sheet region
is a clear evidence for the magnetic annihilation.

\begin{figure}
\resizebox{42mm}{!}{\includegraphics{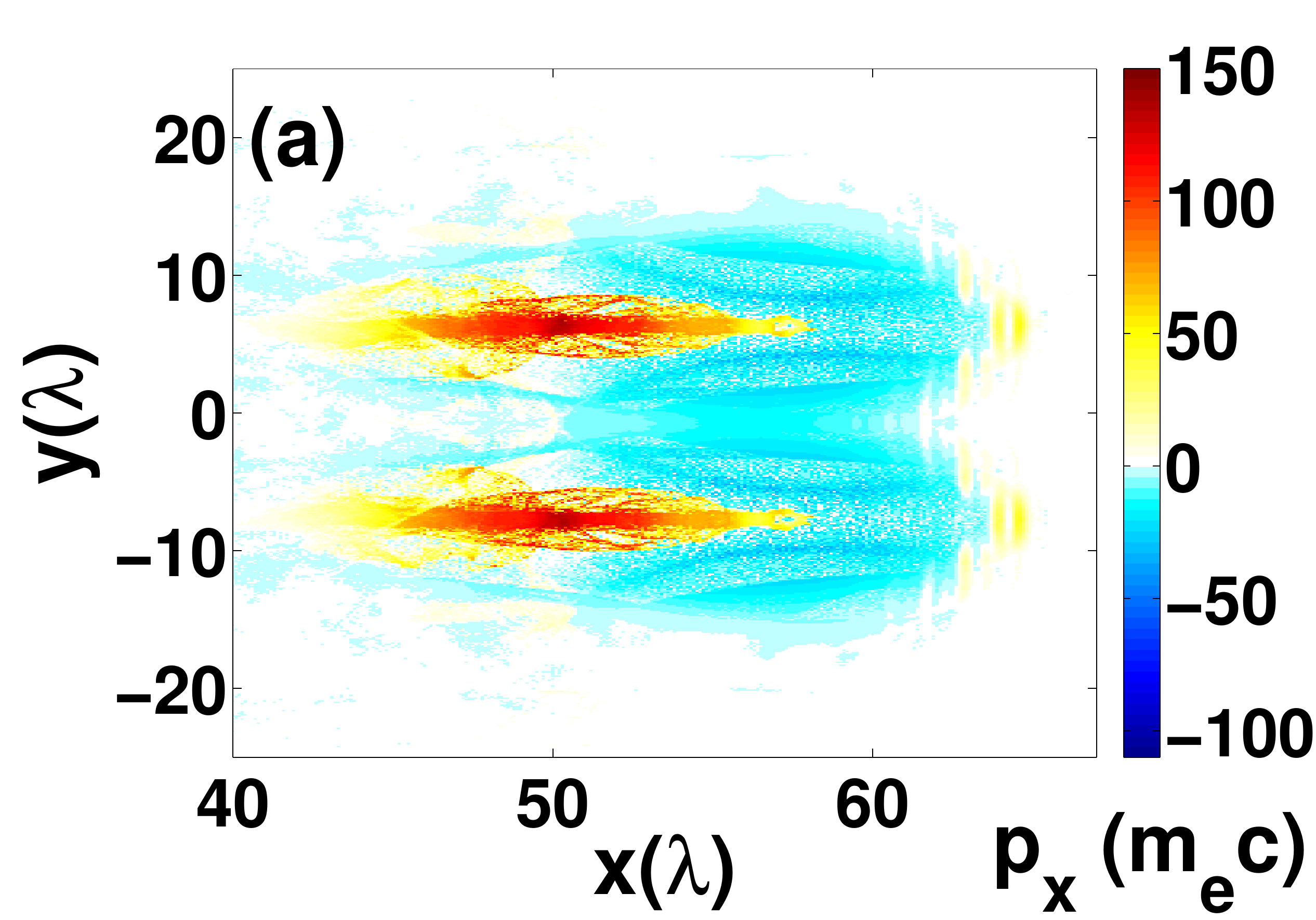}}
\resizebox{42mm}{!}{\includegraphics{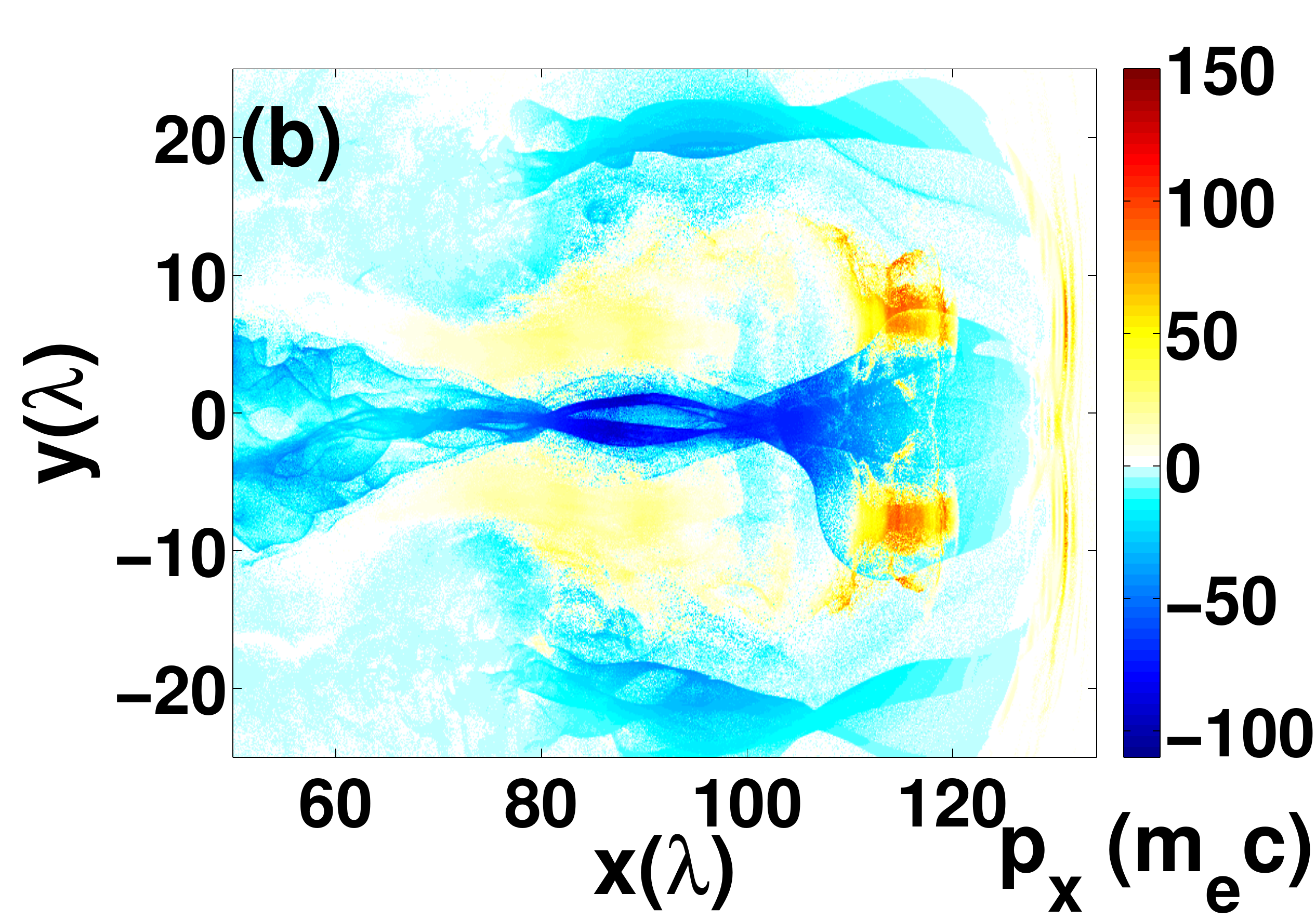}}
\caption{(color online)
The distributions of the electron longitudinal momentum $p_x$
(averaged over one cell) at $69~T_0$ and $138~T_0$ are plotted in (a) and (b), respectively.}
\label{Fig4}
\end{figure}

The inductive electric field can also be used to accelerate efficiently positrons. We
perform a simulation with a mixture plasma of $10\%$ of positrons and $90\%$ of protons.
Positron generation by laser plasma interaction has been demonstrated in Ref. \cite{Positron2010}.
The separation between the two laser pulses is enlarged from $14~\lambda$ to $20~\lambda$.
If the pulse separation is too small, all the positrons encountered by the laser pulses
are pushed forward by the ponderomotive force. As a result,
when the inductive electric field grows,
there are no positrons left in the current sheet.
When the pulse separation is enlarged, a fraction of the positrons in
the current sheet is not pushed forward. They are squeezed together by the transverse ponderomotive force and form a high density region between the two bubbles. With the growth of the inductive electric field, they are accelerated
forward up to high energy and forms a high density positron bullet.
All the other simulation parameters are the same as in the simulations discussed above.
Figure 5(a)
shows the positron density distribution at $171~T_0$.
The high density peak around $x=170~\lambda$ indicates positrons directly accelerated
by the laser ponderomotive force. Behind that, the high energy positron bullet is accelerated by
the inductive electric field. The bullet has maximum momentum equal to $p_x \approx 350~m_e c$, which is much higher
than the positron momentum in the first peak corresponding to the laser direct accelerated positrons (see Fig. 5(b)).
The energy spectrum evolution of the positron bullet is plotted in Fig. 5(c).
The peak energy increases from $20~\mathrm{MeV}$ to $160~\mathrm{MeV}$ in $60~T_0$. As the spectrum shows, the positron bullet is quasi-monoenergetic with the energy
spread of about $\delta E/E \approx 13\%$. The angular divergence approximately equals $14~^\circ$. The bullet contains the charge of $44~\mathrm{pC/\mu m}$. The energy spectrum shows that the inductive field accelerates the positrons to higher energy than is the case for the electrons. This is because the inductive electric field moves forward with the propagating laser pulses continuously accelerating positrons. By contrast, the accelerated electrons move backwards and quickly leave the region of the inductive field.

\begin{figure}
\resizebox{28mm}{!}{\includegraphics{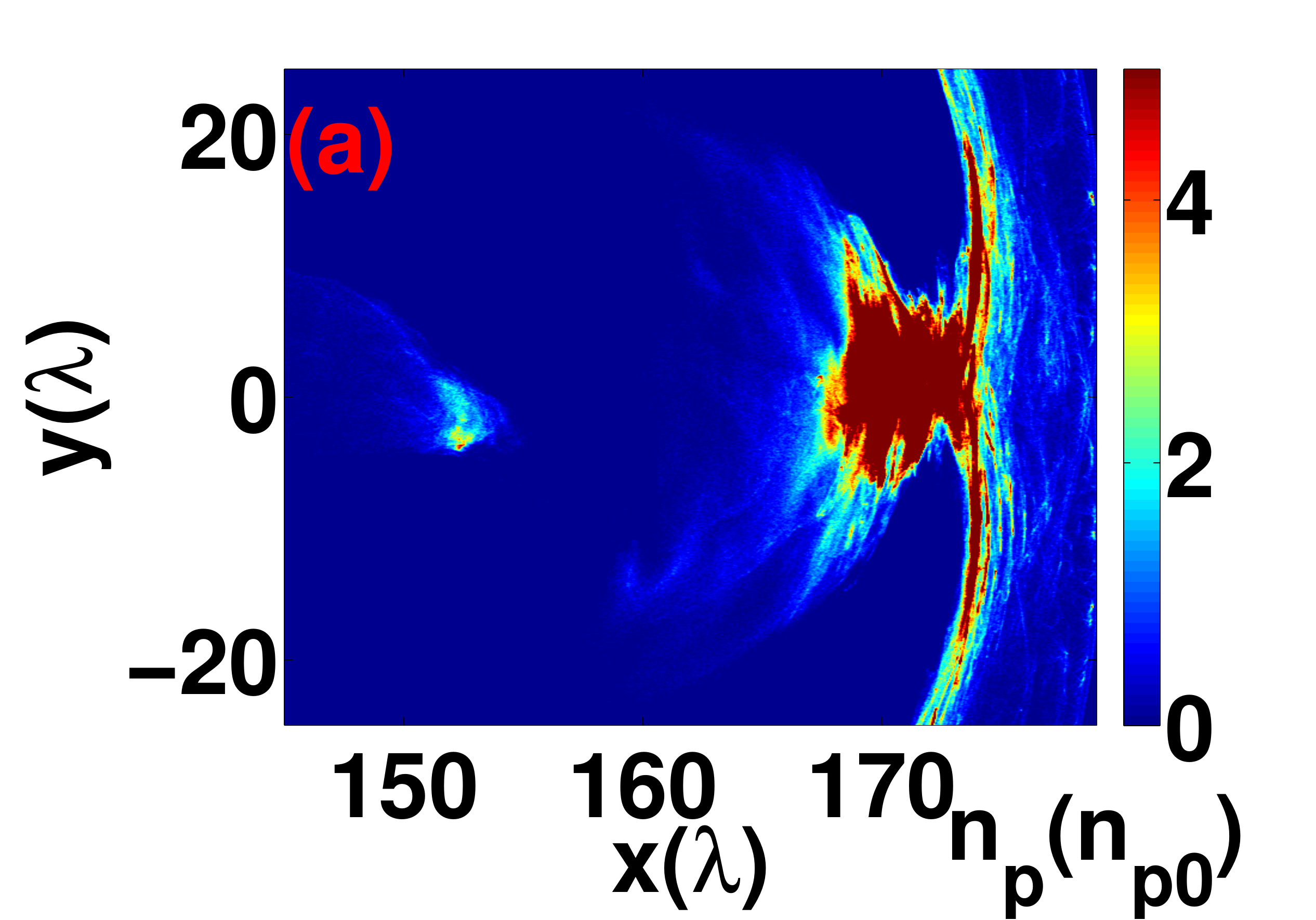}}
\resizebox{28mm}{!}{\includegraphics{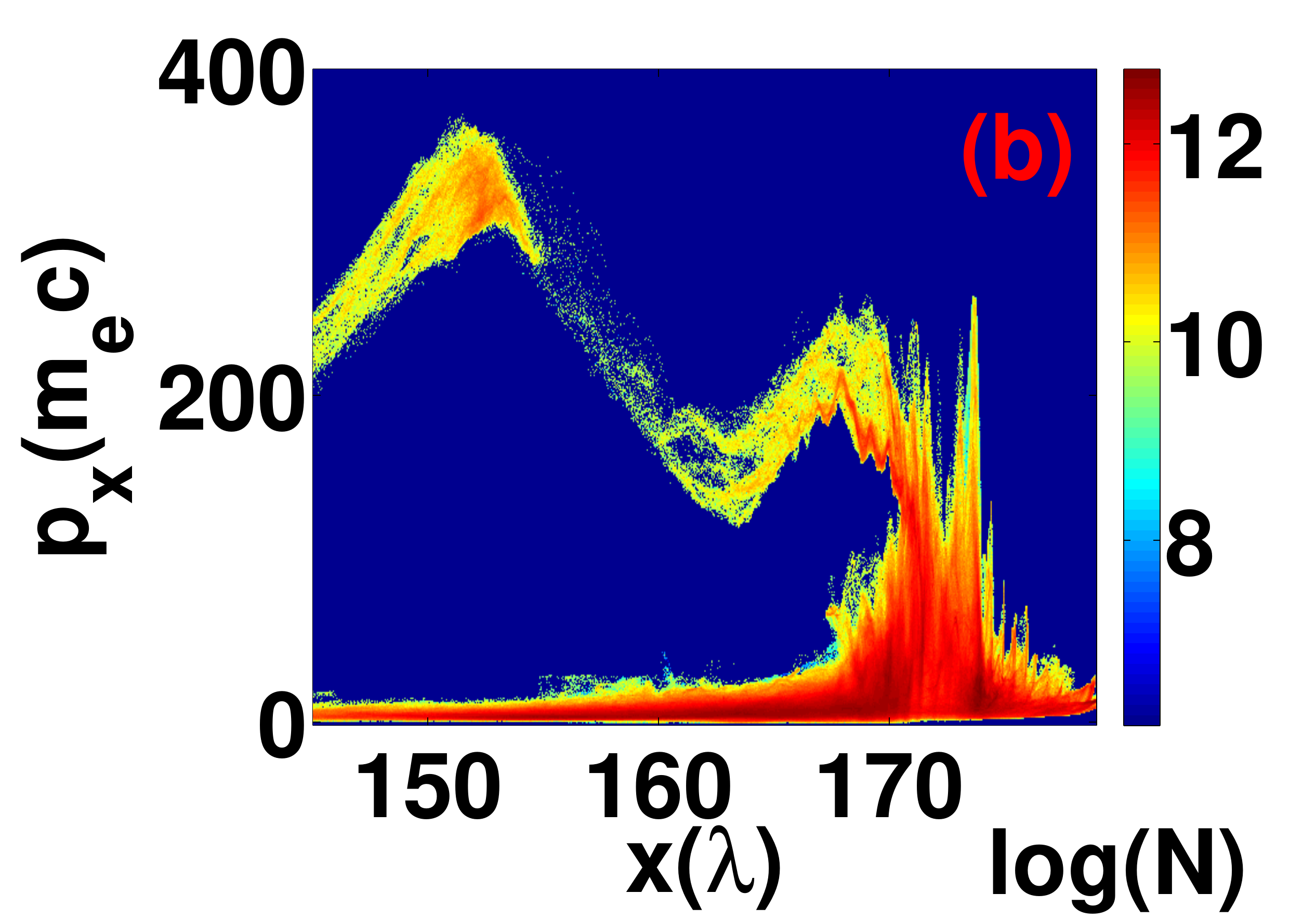}}
\resizebox{28mm}{!}{\includegraphics{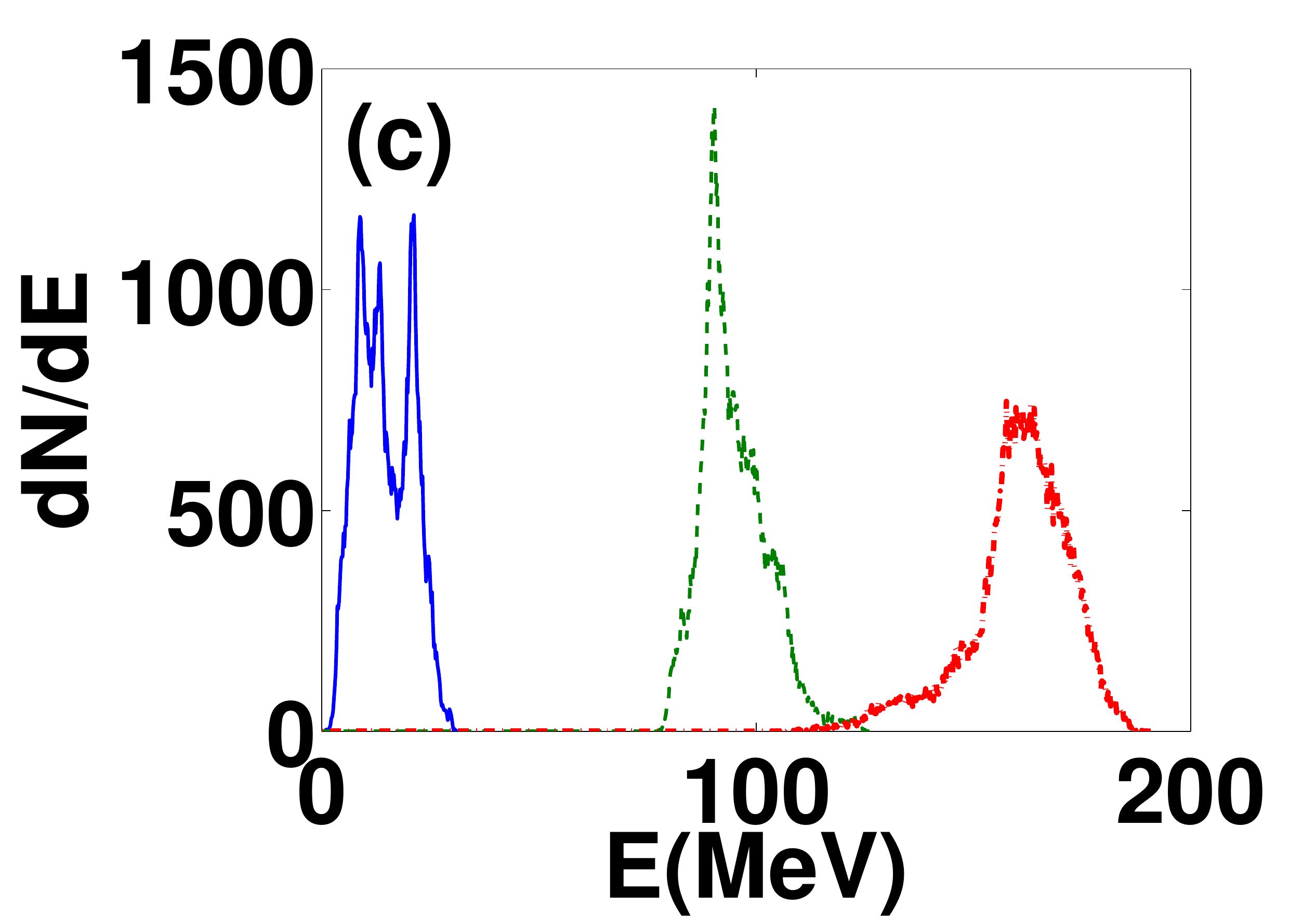}}
\caption{(color online)
The density and momentum $x-p_x$ distribution of positrons at $171~T_0$
are plotted in (a) and (b), respectively. Here $n_{p0} = 0.01~n_c$ is the initial positron density. (c) The energy spectrum of the
inductive field accelerated positron bullet at $120~T_0$ (blue), $150~T_0$ (green) and $180~T_0$ (red).}
\label{Fig5}
\end{figure}

In conclusion, this letter identifies a new regime of collisionless relativistic magnetic annihilation using petawatt lasers operating on very fast time scales. In the magnetic annihilation region, the variation of $\nabla \times \mathbf{B}$ can only be compensated by the displacement current. The associated electric field in the current sheet accelerates electrons to very high energy, opposite to the laser propagation direction. Similarly, the mechanism can efficiently accelerate positrons in the forward direction, which could be used as a signature of this new regime in an experiment. This paper provides predictive simulations for the upcoming petawatt laser installation like ELI \cite{ELIWhitebook}.

\begin{acknowledgments}
This work was supported by ELI (Project No. CZ.1.05/1.1.00/02.0061).
The simulations were performed on the IT4Innovations national supercomputing center
of Czech Republic. The EPOCH code was developed as part of the UK EPSRC funded projects EP/G054940/1.
\end{acknowledgments}


\end{document}